\documentclass[prl, twocolumn, showpacs]{revtex4}
\usepackage{graphics}
\usepackage{amssymb}
\usepackage{amsmath}

\newcommand{\rd}{\mathrm{d}}
\begin{document}


\author{Douglas Armstead}
\email{dna2@physics.umd.edu}
\altaffiliation{\\ Department of Physics}
\altaffiliation{Institute for Research in Electronics and Applied Physics.}
\author{Brian R. Hunt}
\altaffiliation{Department of Mathematics and Institute for Physical
Science and Technology.}
\author{Edward Ott}
\altaffiliation{Department of Physics}
\altaffiliation{Department of Electrical and Computer Engineering}
\altaffiliation{Institute for Research in Electronics and Applied Physics.}
\affiliation{University of Maryland, College Park, Maryland 20742}

\title{Long Time Algebraic Relaxation in Chaotic Billiards}
\date \today

\begin{abstract}
The long time algebraic relaxation process in spatially periodic billiards 
with infinite horizon is shown to display a self-similar time asymptotic form.
This form is identical for a class of such billiards, but can be 
different in an important special case. 
\end{abstract}


\maketitle

\begin{figure}
\begin{center}
\resizebox{80mm}{!}{\includegraphics{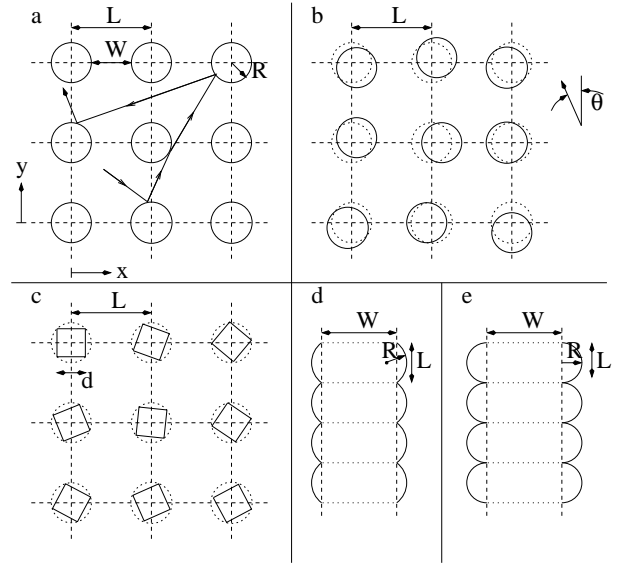}}
\caption{
Infinite horizon billiards.  (a) Circular scatterers of radius $R$ whose 
centers are located on a square grid of periodicity length $L>2R$.  This 
billiard is also referred to as the Sinai billiard \cite{Sinai}.  (b) 
The circular 
scatterers of (a) are each given a random displacement $\vec{\Delta}$ that is 
uniform in the disc $|\vec{\Delta}| \leq \Delta_{0} < (L-2R)/2$.  (c) 
Randomly oriented squares of edge length $d$ where the centers of the squares 
are located on a square grid of periodicity length $L > \sqrt{2} d$. (d) A 
channel whose walls are circular arcs each subtending the same angle which is 
less than $\pi$ radians. (e) Similar to (d) but the circular arcs subtend 
$\pi$ radians (i.e., they are semicircles); orbits in this billiard can be 
mapped to orbits in the stadium billiard of Bunimovich \cite{Bunimovich} by 
reflection at the horizontal dashed lines in the figure.}
\label{fig:examples}
\end{center}
\end{figure}

Diffusion and transport of a population of particles whose motion occurs in an 
infinite periodic array of scatters is a basic problem of continuing interest 
(e.g., Refs. \cite{Geisel} and \cite{Sinai}).  Here we consider billiards, 
i.e., systems in 
which a point particle moves in straight line orbits with constant velocity,
executing specular reflection from fixed boundaries 
(e.g, Refs.~\cite{Sinai, Bunimovich, Bleher, Lee}).
Several examples are shown in Fig.~\ref{fig:examples} and described in the 
figure caption.  These examples all have channels in which a particle 
traveling in the proper direction (either exactly horizontally or vertically 
for the examples in Fig.~\ref{fig:examples}) never experiences reflection; 
i.e., these are \textit{infinite horizon billiards}.  Particle motion without 
reflection occurs only for a zero measure set of initial conditions, but 
it represents a possible source of deviation from classical diffusive 
behavior.  Deviations from classical diffusive behavior (in particular, 
superdiffusive transport) are of great interest in a variety of physical 
situations \cite{Geisel, Afanasiev, ShlesingerBook, Shlesinger, Zaslavsky, 
Zumofen, Ishizaki, Benkadda} and are often associated with the simultaneous 
presence 
of chaotic regions and invariant phase space surfaces [i.e., 
Kolmogorov-Arnold-Moser (KAM) surfaces].  In those cases the anomalous 
transport is associated with the ``stickiness'' \cite{Karney, Hanson, Meiss} 
of the KAM surfaces; 
chaotic particles near these KAM surfaces tend to remain near them for long 
periods of time during which they experience long flights \cite{ShlesingerBook}.  
An analogous 
phenomenon is present in the infinite horizon billiards of 
Fig.~\ref{fig:examples} in that particles in the channels experience long 
flights if their direction of travel is nearly aligned with the channel.  
While KAM stickiness remains a rather difficult phenomenon to fully analyze, 
infinite horizon billiards, like those in Fig.~\ref{fig:examples}, are much 
simpler to study and offer possible insights into the general phenomenon of 
nonclassical diffusive transport. 

Our work on infinite horizon billiards will concentrate on the relaxation  of 
particle distributions to their invariant long-time asymptotic form. In 
particular, we will study how a distribution, initially having no particles 
in phase space regions corresponding to long flights, repopulates these 
regions.  We find, using analytical and numerical methods, that this 
repopulation process occurs in a self-similar manner \cite{diffusion}.  That 
is, at long time 
the relaxing distribution function assumes a particular invariant form when 
expressed in terms of a properly scaled variable.  Furthermore, this 
distribution function is the same for all the billiards in 
Figs.~\ref{fig:examples}(a)-(d), but is different for the billiard in 
Fig.~\ref{fig:examples}(e).  The billiard in Fig.~\ref{fig:examples}(e) is of 
particular interest because its infinite flight invariant set is, in an 
appropriate sense, more sticky than are the invariant sets in the other 
examples.  

It is important to note that the billiards in Fig.~\ref{fig:examples} all 
admit an invariant particle probability distribution function (PDF).  
In particular, considering a monoenergetic ensemble of particles, a 
particle distribution function that is uniform in the accessible area of the 
billiard and isotropic in the angle $\theta$ of the particle velocity vector 
$\vec{v}$ is time invariant.  (Here $\theta$ is defined by 
$|\vec{v} \cdot \vec{y_{o}}|=|\vec{v}|\cos{\theta}$, 
$0 \leq \theta \leq \pi/2$ and $\vec{y_{o}}$ is a unit vector in the vertical
direction.)  Our main concern in this paper is how this invariant 
distribution, isotropic in $\theta$, is approached.

Since the 
channels where infinite flights occur are of most interest, we are primarily 
concerned with the manner in which particles enter and leave the vicinity of 
the vertical infinite flight orbits. For the purpose of exposing the essential 
features of this problem, we consider the relaxation of an initial 
distribution, $P(x,y, \theta, 0)$, that has no particles in the vicinity of 
the infinite flight orbits.  In particular, we take $P(x,y, \theta, 0)$ to be 
zero outside one of the size $L$ cells indicated in 
Fig.~\ref{fig:examples}, while within the accessible part of such a cell

\begin{equation} \label{eq:initDist}
P(\bar{x},\bar{y}, \theta, 0)= \left\{
\begin{array}{ll}
0 & \mathrm{if} \  \theta \leq \theta_{max}, \\
K & \mathrm{if} \  \theta_{max} \leq \theta \leq \pi/2,
\end{array} \right. 
\end{equation}
where $K$ is a normalization constant, $\theta_{max} \ll 1$.
Our main result is that for large time
$\hat{P}(\theta,t) \equiv \int P(x,y, \theta, t) \rd x \rd y$,
where the integral is over $(x, y)$ restricted to the vertical channels,
approaches a scaling form depending on the single variable $\phi=v \theta t/W$.
That is, for large $t$ and $\phi$ bounded,

\begin{equation} \label{eq:selfSim}
\hat{P}(\theta,t) \cong S_{j}(\phi); \ j=0 \ \mathrm{or} \ 1.
\end{equation}
Furthermore, the form of the scaling function on the right hand side of 
(\ref{eq:selfSim}) is the same for the billiards in 
Figs.~\ref{fig:examples}(a)-(d).  We call this function $S_{0}(\phi)$.  For 
the billiard of Fig.~\ref{fig:examples}(e), however, the scaling function,
denoted $S_{1}(\phi)$, takes another form.  In the definition 
$\phi=v\theta t/W$, $W$ is defined as the region of free vertical flights 
[for billiard (a) $W=L-2R$, for billiard (b) $W=L-2(R+\Delta_{0})$, for 
billiard (c) $W$ is the distance between the dashed circles ($W=L-\sqrt{2}d$), 
and for billiards (d) and (e) $W$ is as shown in Fig.~\ref{fig:examples}.]

\begin{figure}
\resizebox{80mm}{!}{\includegraphics{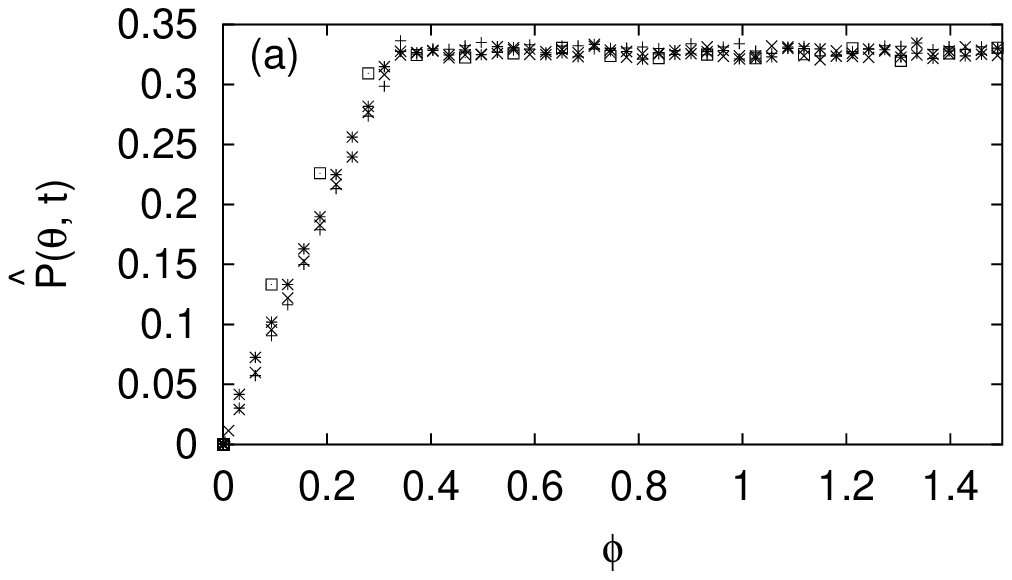}}
\resizebox{80mm}{!}{\includegraphics{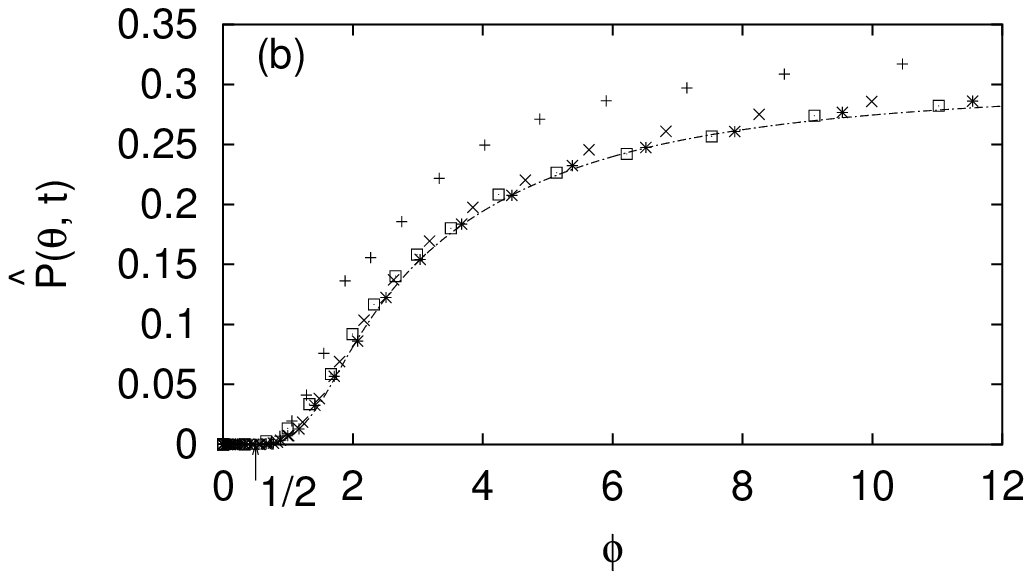}}
\caption{(a) The time evolution of an initially uniform angular distribution 
of particles in the $\theta$ interval $[.3,\pi /2]$ for the billiard of 
Fig.~\ref{fig:examples}(a) with $L/R=3$. The distribution is shown at 
four times; + at $t=9W/v$, $\times$ at $27W/v$, $*$ at $81W/v$, and 
$\square$ at $243W/v$. 
(b) Distribution function for the billiard of Fig.~\ref{fig:examples}(e) with 
$W/R=1$ at the same four times as in (a) and with the same initial $\theta$ 
distribution as in (a).  The solid line in (b) is obtained from the theory, 
Eqs.~(\ref{eq:theory1}) and (\ref{eq:theory2}).
}
\label{fig:data}
\end{figure}

Figure \ref{fig:data}(a) shows histogram approximations to $\hat{P}(\theta, t)$
plotted versus $\phi=v\theta t/W$ for the billiard of 
Fig.~\ref{fig:examples}(a) with $L=3R$ at four times ($+$, $\times$,  $*$, and 
$\Box$ 
correspond to $t=9W/v$, $27W/v$, $81W/v$, and $243W/v$ respectively).
These data are obtained by computing the orbits of a large number of initial 
conditions chosen randomly from the distribution function 
(\ref{eq:initDist}). Results similar to those in Fig.~\ref{fig:data}(a) are 
also obtained for the billiards of Figs.~\ref{fig:examples}(b, c, and d).
In all these cases, at long time, the distribution increases linearly with 
$\phi$ from $\phi=0$, until some critical value $\phi_{0}$ past which the 
distribution is constant in $\phi$,

\begin{equation} \label{eq:S0}
\hat{P}(\theta, t) \cong S_{0}(\phi)= \left \{
\begin{array}{ll}
C \phi & \mathrm{for} \  \phi \leq \phi_{0}, \\
C \phi_{0} & \mathrm{for} \  \phi \geq \phi_{0}.
\end{array} \right. 
\end{equation}
In contrast to the results [e.g., Fig.~\ref{fig:data}(a)] for the billiards of 
Figs.~\ref{fig:examples}(a-d), the billiard of Fig.~\ref{fig:examples}(e) 
gives a very different time asymptotic distribution. This is shown by the
histogram approximations (Fig.~\ref{fig:data}(b)) 
to $\hat{P}(\theta,t)$ obtained for the billiard of 
Fig.~\ref{fig:examples}(e) with $W/R=1$ and the same four times as in 
Fig.~\ref{fig:data}(a).  Note that in this case the long time distribution 
function is identically zero for $\phi < 1/2$.

We now show how the time asymptotic distributions in Figs.~\ref{fig:data} 
arise.  We start with the distribution Eq.~(\ref{eq:S0}) 
[Fig.~\ref{fig:data}(a)] applicable to the billiards in 
Figs.~\ref{fig:examples}(a-d).  As we will subsequently see, for all the 
billiards in Figs.~\ref{fig:examples}(a-d), the scattering of a long flight 
upon reflection from a channel wall leads to a much more drastic change in 
the angle of a particle's velocity vector than in the case of the billiard in 
Fig.~\ref{fig:examples}(e).

\begin{figure}
\resizebox{80mm}{!}{\includegraphics{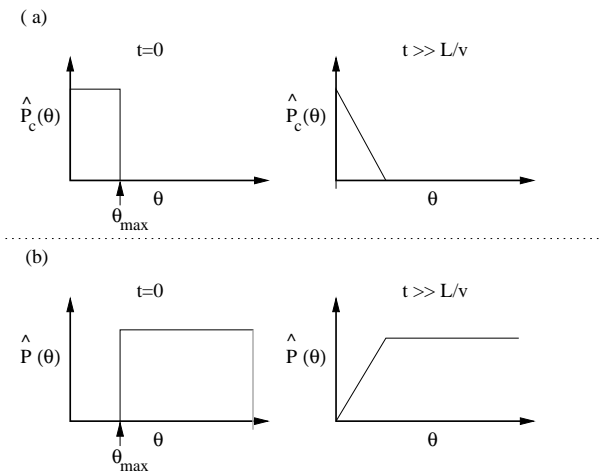}}
\caption{The evolution of $\hat{P}_{c}(\theta, t)$ and
$\hat{P}(\theta,t)$.}
\label{fig:complement}
\end{figure}

For example, in the case of the billiard in Fig.~\ref{fig:examples}(a), the 
angular deflection for small $\theta$ is typically of order $\theta^{1/2}$, 
which is much larger than $\theta$.  In the case of the billiards in 
Fig.~\ref{fig:examples}(b-d), a particle moving nearly parallel to a channel 
axis
is scattered by an angle of order one.  Furthermore, after a large deflection, 
the orientation of the particle's velocity vector is rapidly randomized by a 
succession of many reflections which, since the particle is no longer in a 
long flight, occur in a relatively short time.  These considerations lead
us to a model for the cases in 
Figs.~\ref{fig:examples}(a)-\ref{fig:examples}(d) 
in which we adopt the model hypothesis that, when a particle in a long flight 
suffers a collision with a billiard wall, the orientation of its velocity 
vector is randomly scattered with uniform probability density in $[0,2\pi]$.
We wish to determine the evolution from the initial condition 
(\ref{eq:initDist}) in the case $\theta_{\textrm{max}} \ll 1$.
This initial distribution is equal to the initial distribution 
$P_{b}(\bar{x},\bar{y},\theta,0)=K$ for all $\theta$ minus the initial 
condition $P_{c}(\bar{x},\bar{y},\theta,0)=K$ for $\theta < \theta_{max}$ and 
$0$ otherwise.  The distribution $P_{b}$ remains unchanged when it is 
evolved forward in time (it is an invariant distribution).
Thus, to find the evolution from initial condition (\ref{eq:initDist}), we 
can determine the evolution from $P_{c}$, and then subtract it from $K$.
The long time evolution from $P_{c}$ can be found by considering the time at 
which particles are scattered.  Consider, for example, the billiard of
Fig.~\ref{fig:examples}(d), and a particle with a small initial $\theta_{o}$.
Suppose the particle is located in the channel at a distance $\Delta x$
from the boundary of the channel with which it will collide [left or right
vertical dashed line in Fig.~\ref{fig:examples}(d)].  If 
$\Delta x < vt \sin \theta_{o} \cong vt\theta_{o}$, the particle
scatters before time $t$; if $\Delta x > vt\theta_{o}$, it does not scatter.  
For particles in the channel, $\Delta x <W$, so every
particle with $\theta_{o}>W/(vt)$ must have scattered at least once.
We assume that $t >W/(v\theta_{max}) \equiv t_{o}$.  Since 
$\theta_{\textrm{max}}$ is small, the scattered particles contribute a
small positive value of order $\theta_{\textrm{max}}$ to 
$\hat{P}_{c}(\theta,t)$ in $\theta$.  Thus 
$\hat{P}_{c}(\theta,t)$ is small (i.e., of order $\theta_{\textrm{max}}$)
for $\theta > W/vt$.  For $\theta_{o}< \Delta x/(vt)$, $t>t_{o}$, the 
particle has not yet scattered.  Assuming that the initial spatial 
distribution of particles in the channel is uniform, the fraction of 
particles with initial angle $\theta_{o}$ that have scattered is 
$\theta_{o}vt/W$.  Thus

\begin{equation}
\hat{P}_{c}(\theta,t) \cong \left\{
\begin{array}{ll}
K(1-\theta vt/W) & \textrm{ for } \theta< W/(vt),\\
0 & \textrm{ for } \theta > W/(vt),
\end{array} \right.
\end{equation}
where we have neglected the small, order $\theta_{max}$, contribution to 
$\hat{P}_{c}(\theta,t)$ from scattered particles.  Subtracting $\hat{P}_{c}$
from $\hat{P}_{b}$ as illustrated in Fig.~\ref{fig:complement}, we obtain the 
time asymptotic form in Eq.~(\ref{eq:S0}).

\begin{figure}
\resizebox{80mm}{!}{\includegraphics{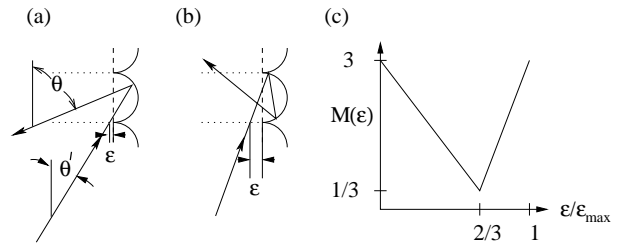}}
\caption{(a) A one-bounce encounter with the wall resulting in the reversal 
of the particle's vertical velocity component. (b) A two-bounce encounter 
with the wall resulting in an unchanged direction of the particle's vertical 
velocity component. (c) $M(\epsilon)$ verses $\epsilon$ for 
$\epsilon_{max}=2R \tan  \theta' \ll 1$.
}
\label{fig:bounce}
\end{figure}

We now consider the evolution of initial condition (\ref{eq:initDist}) in the
billiard of Fig.~\ref{fig:examples}(e) for orbits experiencing long flights,
Fig.~\ref{fig:data}(b).  When an orbit in a long flight encounters the channel 
wall it may experience either one reflection (Fig.~\ref{fig:bounce}(a)) or two 
reflections (Fig.~\ref{fig:bounce}(b)).  A geometrical analysis \cite{Lee} 
shows that 
for $\theta' \ll 1$, the relation between the angle $\theta'$ before the 
encounter and the angle $\theta$ after the encounter is 
$\theta \cong M(\epsilon)\theta'$, where $\epsilon$ is the distance illustrated in 
Figs.~\ref{fig:bounce}(a,b) and $M(\epsilon)$ is the piecewise linear 
function shown in Fig.~\ref{fig:bounce}(c).  Note from Fig.~\ref{fig:bounce}(c)
that the trajectory angle can increase or decrease at most by a factor of $3$ 
upon an encounter with the wall.  Thus a long flight is again followed by a 
long flight, in contrast to the billiards of Figs.~\ref{fig:examples}(a-d).
That is, the infinite flight invariant set for the billiard of 
Fig.~\ref{fig:examples}(e) is more sticky than are those of the other 
billiards, Figs.~\ref{fig:examples}(a-d).  Note also that, for $\theta \ll 1$,
 $\epsilon$ at the next encounter with the end of the channel 
is very sensitive to a small percentage change in $\theta$.  This
fact, together with the form of $M(\epsilon)$ shown in 
Fig.~\ref{fig:bounce}(c), leads us to adopt a model for our subsequent 
analysis whereby upon an encounter with the wall $\theta' \to \theta=M\theta'$ 
where $M$ is chosen randomly with uniform probability density in the interval
$(1/3, 3)$.  As previously noted $S_{1}(\phi) \equiv 0$ for  $\phi < 1/2$.  
This can be understood as follows.  $\hat{P}(\theta, 0)=0$ for 
$\theta < \theta_{0}$, and on subsequent wall encounters $\theta$ can decrease
at most by a factor of $3$.  Therefore, after $n$ encounters the minimum angle
is $\theta_{n}=\theta_{0}/3^{n}$ and the minimum time for these $n$ encounters
is 
$t_{n}= \sum_{n'=0}^{n-1} W/(v \theta_{n'}) \approx (W/v)3^{n}/(2\theta_{0})$ 
for $n \gg 1$, yielding a minimum value of $1/2$ for $\phi$.

Let $Q(\theta, t) \rd \theta$ be the rate at which particles 
are scattered into the channel with angle between $\theta$ and
$ \theta + \rd \theta$.  Particles that enter the channel with small angle 
$\theta$ remain in the channel for the time 
$W/(v \sin \theta) \cong W/(v \theta)$.  Then for large $t$ and small $\theta$,
since the number of particles in the interval $\theta$ to $\theta+\rd \theta$ 
at time $t$ is approximately $S_{1}(\phi) \rd \theta$, we have
$S_{1}(\phi)= \int_{t-W/(v\theta)}^{t} Q(\theta, t') \rd t'$.  
Since the left hand side is only a function of $\phi$, $Q(\theta, t)$ must 
have the asymptotic scaling form $Q(\theta, t)=(v \theta/W)C(\phi)$.  Thus 

\begin{equation} \label{eq:theory1}
S_{1}(\phi)= \int_{\phi-1}^{\phi} C(\phi) \rd \phi.
\end{equation}
To obtain an equation for $C(\phi)$ consider a particle in a long flight that,
at time $t'$, has just suffered an encounter with a channel wall.  Denote 
its angle as a result of this encounter by $\theta' \ll 1$ (see 
Figs.~\ref{fig:bounce}(a,b)).  The particle then experiences a long flight 
of duration $W/(v\theta')$ before it next encounters a channel wall.  
According to our model, the velocity angle $\theta$ just after this encounter
at time $t$ is $\theta= M \theta'$ where $M$ is random with uniform 
probability density in the interval $(1/3, 3)$.  During a time interval $t$ to 
$t +\rd t$ the number of particles scattered into the interval $\theta$ to 
$\theta +\rd \theta$ is 
$Q(\theta, t) \rd \theta \rd t = 3/8 \int_{1/3}^{3} 
\left [ Q( \theta', t') \rd \theta' \rd t' \right ] \rd M$,
where $\theta'=\theta/M$, $t'=t- \frac{W}{v\theta'}$, and $3/8=1/(3-1/3)$ is 
the probability density of the random variable $M$.  Now substituting 
$Q(\theta, t)=(v\theta/W)C(\phi)$ and making a change of the integration 
variable $m=1/M$, we arrive at the following integral equation for $C(\phi)$, 

\begin{equation} \label{eq:theory2}
C(\phi)=3/8 \int_{1/3}^{3} C(m\phi -1) \rd m \equiv \mathcal{L}[C(\phi)]
\end{equation}
whose solution along with (\ref{eq:theory1}) determines $S_{1}$.  We 
numerically solve (\ref{eq:theory2}) by iteration of 
$C_{i+1}(\phi)= \mathcal{L}[C_{i}(\phi)]$ starting with $C_{0}(\phi)=0$ if 
$\phi < 1$ and $C_{0}(\phi)=1$ if $\phi \geq 1$.  We use $10^{4}$ 
grid points uniformly spaced in the $\phi$ interval $[0, 50]$ with 
$C_{i}(\phi)$ set to one for $\phi > 50$ (this corresponds to the boundary 
condition $C(\phi) \to \mathrm{const.}$ for $\phi \to \infty$).  The results
are virtually unchanged if the $\phi$ interval is increased to $[0, 100]$ or 
if the 
number of grid points is increased, and virtually identical results (i.e., 
convergence) were found for several $i > 12$.  A result obtained from such a 
calculation is shown as the solid curve in Fig.~\ref{fig:data} $(i=15)$ and is 
in good agreement with our numerical histograms computed from many particle 
orbits \cite{AHO1}.

In conclusion, we have shown that the long-time repopulation of long flights 
in infinite horizon billiards proceeds in a self-similar manner 
[Eq.~(\ref{eq:selfSim})].  The corresponding relaxation of the distribution 
[which follows from (\ref{eq:selfSim})] is algebraic rather than exponential;
e.g., the $\theta$ integral of $|\hat{P}(\theta,t)-\hat{P}(\theta, \infty)|$
decays like $1/t$.
Furthermore, the self-similar forms found for a class of billiards 
that include those of Figs.~\ref{fig:examples}(a-d) are the same, 
Eq.~(\ref{eq:S0}).  On the other hand, the special case of the billiard of 
Fig.~\ref{fig:examples}(e) is shown to yield a different time-asymptotic 
self-similar form, and this may be ascribed to the greater stickiness of the 
invariant set of this billiard.

We thank J. R. Dorfman for discussions.
This work was supported by the Office of Naval Research (Physics) and by the 
National Science Foundation (Grants PHYS098632, DMS010487).

\end{document}